\documentclass[11pt]{article}
	
	\newcommand{\blind}{0}
	
    \makeatletter
    \renewcommand\section{\@startsection {section}{1}{\z@}%
                                       {-3.5ex \@plus -1ex \@minus -.2ex}%
                                       {2.3ex \@plus.2ex}%
                                       {\normalfont\fontfamily{phv}\fontsize{16}{19}\bfseries}}
    \renewcommand\subsection{\@startsection{subsection}{2}{\z@}%
                                         {-3.25ex\@plus -1ex \@minus -.2ex}%
                                         {1.5ex \@plus .2ex}%
                                         {\normalfont\fontfamily{phv}\fontsize{14}{17}\bfseries}}
    \renewcommand\subsubsection{\@startsection{subsubsection}{3}{\z@}%
                                        {-3.25ex\@plus -1ex \@minus -.2ex}%
                                         {1.5ex \@plus .2ex}%
                                         {\normalfont\normalsize\fontfamily{phv}\fontsize{14}{17}\selectfont}}
    \makeatother
	
\usepackage{amsmath,amssymb}
\usepackage{subcaption}
\usepackage{authblk}
\usepackage{float}
\usepackage{graphicx}
\usepackage{enumerate}
\usepackage{algpseudocode}
\usepackage{algorithm}
\usepackage{booktabs,tabularx}
\usepackage{tabularx}
\usepackage{booktabs}
\usepackage{xcolor}
\usepackage[sort, numbers]{natbib}
\usepackage{url} 
	
\begin{document}
		
		\def\spacingset#1{\renewcommand{\baselinestretch}%
			{#1}\small\normalsize} \spacingset{1}
\if0\blind
{
    \title{Hierarchical Semi-Markov Models with Duration-Aware Dynamics for Activity Sequences}
    \author[1]{Rohit Dube}
    \author[2]{Natarajan Gautam}
    \author[1]{Amarnath Banerjee}
    \author[3]{Harsha Nagarajan}

\affil[1]{Industrial and Systems Engineering, Texas A\&M University, College Station, USA}
\affil[2]{Electrical Engineering and Computer Science, Syracuse University, NY, USA}
\affil[3]{Applied Mathematics and Plasma Physics, Los Alamos National Laboratory, Los Alamos, USA}
    
    \date{}
    \maketitle
} \fi

\if1\blind
{
    \title{Interval Prediction of Electricity Demand Using a Cluster-Based Block Bootstrapping Method}
    \author{Author information is purposely removed for double-blind review}
    \maketitle
} \fi

		
\begin{abstract}
Residential electricity demand at granular scales is driven by what people do and for how long. Accurately forecasting this demand for applications like microgrid management and demand response therefore requires generative models that can produce realistic daily activity sequences, capturing both the timing and duration of human behavior. This paper develops a generative model of human activity sequences using nationally representative time-use diaries at a 10-minute resolution. We use this model to quantify which demographic factors are most critical for improving predictive performance.

We propose a hierarchical semi-Markov framework that addresses two key modeling challenges. First, a time-inhomogeneous Markov \emph{router} learns the patterns of ``which activity comes next." Second, a semi-Markov \emph{hazard} component explicitly models activity durations, capturing ``how long" activities realistically last. To ensure statistical stability when data are sparse, the model pools information across related demographic groups and time blocks. The entire framework is trained and evaluated using survey design weights to ensure our findings are representative of the U.S. population.

On a held-out test set, we demonstrate that explicitly modeling durations with the hazard component provides a substantial and statistically significant improvement over purely Markovian models. Furthermore, our analysis reveals a clear hierarchy of demographic factors: Sex, Day-Type, and Household Size provide the largest predictive gains, while Region and Season, though important for energy calculations, contribute little to predicting the activity sequence itself. The result is an interpretable and robust generator of synthetic activity traces, providing a high-fidelity foundation for downstream energy systems modeling.
\end{abstract}

\noindent%

\spacingset{1.5} 
\newpage
\section{Introduction}\label{s:intro}

The use of distributed energy systems, electrification of transportation with the increasing reliance on renewables are transforming the electric power grid. At the granular scales of buildings, neighborhoods, and microgrids, traditional top-down load forecasting models are becoming increasingly inadequate. Demand at the end-user level is not a smooth, aggregate signal but a highly volatile one, driven directly by the stochastic nature of human behavior; when people wake up, work, cook, or charge their electric vehicles \cite{muratori2018impact, Johnson2017Electrical-end-useSystems, 7887751, Saldanha2012MeasuredResolution, Yuan2020ResidentialGrids}. Accurate planning and control of these edge systems, particularly for applications like demand response and local energy market design, therefore necessitate a deeper, more fundamental understanding of their primary driver: \emph{human activity}.

To address this, a significant body of research has focused on ``bottom-up" activity-based models. These works simulate residential electricity demand by first simulating what people are doing, by using Markov modelling in  \cite{capasso2002bottom,Muratori2013ADemand} and bootstrap sampling in \cite{Chiou2011AMethod}. A successful model in this domain must overcome three central challenges: 1) realistically predicting the \textit{sequence} of activities (what comes next?); 2) accurately capturing their \textit{duration} (for how long?); 3) maintaining statistical stability even when segmenting the population by demographic factors to build Markov matrices, which often leads to severe \textit{data sparsity}.

This paper develops a unified framework that simultaneously addresses all three challenges. We propose a hierarchical semi-Markov hazard model that synthesizes a time-inhomogeneous Markov router for sequencing, a hazard-based component for durations, and a hierarchical structure for managing data sparsity while building Markov matrices. Using this model, we conduct a systematic, data-driven evaluation to quantify the effects of hazard and which demographic factors are most critical for predicting activity sequences. 

\subsection{Related Work}\label{s:related_work}
This review covers three strands that frame the study: activity-based residential load modeling (from occupancy-only and Markov simulators to duration-aware methods), time-use diary datasets with emphasis on American Time Use Survey (ATUS) and survey weighting at 10-minute resolution, and hierarchical modeling to stabilize estimates under sparsity. Subsections \ref{subsec: ABRLM} to \ref{subsec:hie model} follow this sequence.

Against this background, the review highlights gaps the present work addresses: understated dwell-time structure in Markov-only models, loss from coarse temporal aggregation, and instability without pooling. These themes connect directly to the research questions on explicit duration modeling and the marginal value of demographic covariates, setting up the modeling Section \ref{s:model} and results Section \ref{s:results}.

\subsubsection{Activity-Based Residential Load Modeling}\label{subsec: ABRLM}
Early work in residential load modeling developed bottom-up frameworks \cite{prina2020classification,zhao2024simulating} linking occupant activities with appliance demand. A seminal study in \cite{capasso2002bottom} proposed one of the first stochastic household load models, simulating appliance usage through Monte Carlo methods informed by occupant behavior. This established the foundation for treating human activity as the driver of residential demand. Later, \cite{Richardson2008ASimulations} advanced this line by introducing the models that applied Markov chains to time-use statistics to generate occupancy and activity sequences, then mapped these sequences to appliance use. Their approach demonstrated that even simple Markov structures could capture the diurnal patterns of household demand.

Parallel work by \cite{Widen2009ConstructingValidation, Widen2010ADemand} integrated time-use diaries with bottom-up appliance models, particularly for domestic lighting. These studies showed that demand models grounded in activity sequences not only reproduced realistic load curves but also aligned well with measured data. However, as several authors noted, purely Markovian simulators and smaller datasets tend to underestimate the persistence of long events, leading to unrealistic distributions of very short or very long activities. 

To address this shortcoming, hybrid models introduced explicit duration components. \cite{page2008generalised} developed an occupant presence model combining Markov chains with survival analysis to capture how long occupants are present or absent. This duration-aware approach provided more realistic occupancy sequences than memoryless models. Similarly, hazard-based sub-models by \cite{HOU2020107126, AHMED2023113558, d2019critical} have been incorporated in later occupancy research to determine dwell times for room presence or appliance use. 

While such hybrid presence models are semi-Markov in spirit—using survival analysis to relax geometric dwell times—they focus on a \emph{binary} presence/absence process and typically assume time-homogeneous or coarsely piecewise-homogeneous dynamics. They do not learn a time-slot–specific destination distribution over multiple activities. Our work directly builds on these insights: we adopt a semi-Markov formulation with an explicit hazard component, ensuring that both \emph{which activity occurs next} and \emph{how long it persists} are modeled not just for occupancy but for various other activities.

Another limitation was often a direct consequence of the smaller datasets available at the time, which constrained the ability to define more detailed activity categories or capture accurate estimates of transition probabilities. In fact, \cite{Widen2010ADemand} explicitly noted that a ``larger and more detailed data set'' covering a longer time period would be a key refinement to improve model fidelity. Our work directly addresses this call by leveraging over two decades of the ATUS data (2003--2024), providing the statistical power needed not only to model durations more accurately but also to robustly estimate parameters across numerous demographic subgroups. The full set of ATUS data variables used in our analysis is listed in Appendix \ref{app:data cols}.

\subsubsection{Time-Use Data and Stochastic Activity Models}\label{subsec: TUS}
Time-use diary surveys (TUS) \cite{BLS_ATUS_microdata_Year,UKTUS_2014_2015_UKDS_8128, BTUS13_Statbel_2013} and electrical end-use datasets \cite{ElectricINL, 2020NorthwestMetering} have become a central resource for energy-oriented activity modeling. The ability of TUS to represent population-level variability in demand timing has made them especially valuable for demand-side management research.

The ATUS \cite{BLS_ATUS_microdata_Year} has been particularly influential. The work done by \cite{Muratori2013ADemand} demonstrated how ATUS data can calibrate Markov-chain activity simulators to produce detailed household demand estimates. Building on this, \cite{mitra2020typical} derived typical U.S. occupancy schedules from ATUS records. In related work, \cite{mitra2021cluster} used clustering to identify dominant archetypes of daily occupancy. Likewise \cite{mitra2021variation} applied clustering methods to ATUS to derive occupancy profiles for residential building simulations.

These studies reinforce two key premises of our approach: (i) large-scale diary data yield better representative activity sequences, and (ii) fine-grained temporal resolution is critical. Unlike prior work that often reduces diaries to larger intervals and binary occupancy schedules, our framework preserves the full 10-minute resolution of multiple ATUS activities and not just presence and absence of the respondents. This enables mapping of specific activities (e.g., cooking, laundry, leisure) to end-use loads in downstream applications.

A further methodological contribution of our work lies in the rigorous use of \emph{survey design weights}. While being a best practice in social science, weighting has often been neglected in energy studies, where diaries are treated as simple random samples. We incorporate ATUS weights throughout model training and evaluation, ensuring that estimated transition probabilities and likelihood comparisons reflect the U.S. population. To our knowledge, this population-weighted stochastic modeling is novel in the context of residential load forecasting.

\subsubsection{Hierarchical Modeling and Demographic Factors}\label{subsec:hie model}
Modeling activity sequences with multiple covariates such as Day-Type, employment, or region poses challenges of sparsity. Stratifying by many attributes quickly fragments the data, reducing the reliability of estimates. In \cite{gelman2007data} it was emphasized that hierarchical pooling can mitigate this by borrowing strength across groups while retaining group-specific deviations.

In the energy domain, clustering has served as an implicit form of pooling. For instance, \cite{aerts2014method} applied hierarchical clustering to Belgian TUS data \cite{BTUS13_Statbel_2013}, identifying a small set of representative occupancy sequences to stabilize building simulations. Our approach differs in that we retain the full fine-grained resolution of activities and instead shrink slot-level transition probabilities (10-minute resolution) toward block-level priors (6-hour resolution), producing stable yet interpretable daily trajectories even in sparse subgroups.

Demographic and contextual drivers of residential demand have also been studied extensively. The contrast between weekdays and weekends is well documented \cite{mitra2020typical, mitra2019defining}, and employment status has been shown to shift energy use toward evenings \cite{lHorincz2021impact}. Gender and number of occupants have produced mixed findings: systematic differences were found in \cite{suomalainen2019detailed} for time at home between men and women but with high within-group variability. Seasonal and regional effects, in contrast, appear to affect the intensity of activities (e.g., heating/cooling loads) rather than the sequence of behaviors themselves \cite{anderson2018explaining,Widen2010ADemand}.

\subsection{Contributions}
The literature on activity-based load modeling spans bottom-up stochastic models, duration-aware hazard methods, and the integration of large-scale time-use diaries. Our work synthesizes these strands by combining (i) a hierarchical semi-Markov framework, (ii) explicit hazard modeling of durations, and (iii) survey-weighted estimation on nationally representative ATUS data. This integration addresses gaps between small-scale, high-fidelity behavioral studies and the coarse scheduling assumptions often used in engineering practice. By situating our contributions within established research while extending them with methodological innovations, we demonstrate how realistic, population-representative activity sequences can form the foundation for more accurate downstream energy system applications.

This paper develops an interpretable modeling framework that directly addresses these gaps. Our primary contributions are:
\begin{itemize}
    \item \textbf{A Unified Hierarchical Framework:} We develop an integrated semi-Markov model that simultaneously captures activity sequencing and durations while ensuring statistical stability through hierarchical shrinkage, even in data-sparse subgroups.
    \item \textbf{Quantifying the Value of Durations:} We rigorously demonstrate, using a bootstrap analysis, that explicitly modeling durations provides a substantial and statistically significant improvement in predictive accuracy over purely Markovian approaches.
    \item \textbf{Identifying Key Demographic Drivers:} We systematically measure the predictive value of individual demographic covariates, revealing a clear hierarchy in which some factors are critical for predicting daily activity patterns.
    \item \textbf{A Robust Generative Tool:} The resulting model is an interpretable and computationally efficient generator of realistic, demographically-conditioned activity sequences, providing a high-fidelity foundation for downstream energy systems modeling.
\end{itemize}

The remainder of this paper is organized as follows. Section \ref{s:data} describes our data source, the ATUS, the preprocessing steps used to construct our modeling dataset, and the experiment design. Section \ref{s:model} provides a detailed walkthrough of the hierarchical semi-Markov model. Section \ref{s:results} presents the empirical results of our model comparisons, followed by a discussion of their interpretation and implications in Section \ref{s:discussion}. Finally, Section \ref{s:conclusion} concludes the paper and suggests directions for future research.

\section{Data and Preprocessing}\label{s:data}

The foundation of our model is the ATUS, a nationally representative diary study conducted by the U.S. Census Bureau. Each diary entry details a 24-hour sequence of activities for a respondent aged 15 or older, beginning at 4 a.m. The raw data provides minute-by-minute episodes with 6-digit activity codes, along with rich demographic information and crucial survey design weights that enable population-level inference.

Our data preprocessing pipeline transforms this raw, event-based data into a structured, fixed-grid format suitable for our models. This involves three key steps: data assembly, temporal discretization, and the creation of a purpose-built activity taxonomy described in the following sections.

\subsection{Data Assembly and Temporal Discretization}
First, we merge several raw ATUS files, including the respondent-level, activity-level, and household roster files (\cite{BLS_ATUS_microdata_Year}), to create a unified dataset for each diary entry. This links each activity episode with the full set of respondent and household characteristics.

While the raw diaries have a minute-by-minute resolution, we discretize each 24-hour diary into a categorical time series of \(T=144\) fixed 10-minute slots. For each slot \(t\), we identify the primary activity performed by respondent \(i\). This process results in a sequence where the activity in each slot is still represented by one of several hundred raw ATUS codes. To create a tractable and interpretable model, these fine-grained codes must be mapped to a smaller, coherent set of states.

\subsection{Activity Taxonomy and Rule-Based Mapping} \label{ss:taxonomy}
To make our model interpretable and focused on energy-relevant behaviors, we developed the 14-state taxonomy shown in Table~\ref{tab:taxonomy}. The design is guided by separating activities based on location (at home vs. out of home) and their likely electricity consumption.

\begin{table}[h!]
\centering
\caption{The 14-State Activity Taxonomy for Modeling.}
\label{tab:taxonomy}
\begin{tabularx}{\textwidth}{@{} l l X @{}} 
\toprule
\textbf{Major Category} & \textbf{Model State (\(\mathcal{S}\))} & \textbf{Description \& Examples} \\
\midrule
\textit{\parbox[t]{4cm}{At-Home,\\(Electric)}} & \textsc{Cooking} & Meal preparation using stove, oven, or microwave. \\
& \textsc{Dishwashing} & Manual or machine dishwashing and kitchen cleanup. \\
& \textsc{Laundry/Ironing} & Using a washer, dryer, or iron. \\
& \textsc{Electric Cleaning} & Using vacuum cleaners, floor polishers, etc. \\
& \textsc{Screens/Leisure} & Watching TV, using computers or game consoles, phone calls. \\
& \textsc{Admin on Devices} & Household finances, email, using printers or scanners. \\
& \textsc{Electric Appliance} & Using power tools, electric motors, etc. \\
\midrule
\textit{\parbox[t]{4cm}{At-Home,\\(Non Electric)}} & \textsc{Sleep} & Includes sleeping and napping. \\
& \textsc{Eating/Drinking} & Consuming meals without active preparation. \\
& \textsc{Personal Care} & Bathing, grooming, dressing. \\
& \textsc{Care at Home} & Childcare or providing help to other household adults. \\
& \textsc{Quiet/Social} & Reading, relaxing, talking, or other quiet activities. \\
& \textsc{Exercise (No Machine)} & Yoga, calisthenics, stretching, etc. \\
\midrule
\textit{Out-of-Home} & \textsc{Out-of-Home} & Encompasses work, shopping, travel, and all other non-home activities. \\
\bottomrule
\end{tabularx}
\end{table}

Mapping the hundreds of raw ATUS codes to our 14 model states is performed using a hierarchical, rule-based procedure. The 6-digit ATUS codes are structured, where the first two digits represent a major activity category (e.g., `01' for Personal Care), the next two define a sub-category, and the final two provide fine-grained detail. Our mapping logic leverages this structure to ensure classifications are transparent and consistent. The rules are applied with the following precedence:

\begin{enumerate}
    \item \textbf{Exact 6-Digit Code Overrides:} We first manually check for specific codes where the energy implication is unambiguous and might otherwise be misclassified. For instance, `020101' (``Interior cleaning") is explicitly mapped to \textsc{Electric Cleaning}.
    \item \textbf{Prefix Rules:} We then use 2- and 4-digit prefixes to classify entire families of activities. This is our primary method for leveraging the ATUS code structure. For example, all codes beginning with the 2-digit prefix `01' (Personal Care) are mapped to the \textsc{Personal Care} state, while codes starting with `0203' are mapped to \textsc{Laundry/Ironing}.
    \item \textbf{Keyword Matching:} For codes not captured by prefixes, we search the activity's text description for specific keywords (e.g., ``cook", ``oven", ``microwave") to assign it to a state like \textsc{Cooking}.
    \item \textbf{Fallback Assignment:} Finally, any remaining unclassified codes are assigned based on coarse, major-category heuristics or default to the safest assignment, \textsc{Out-of-Home}.
\end{enumerate}
This structured mapping process produces a clean, consistent time series \((S_{i,1}, \dots, S_{i,144})\) for each respondent \(i\), which forms the core data for training and evaluating our model.

\subsection{Empirical Motivation for Model Choices}
The selection of a time-inhomogeneous semi-Markov model is not merely a theoretical choice but is directly motivated by strong empirical patterns observed in the ATUS data.

\begin{figure}[!htbp]
\centering
\includegraphics[width=0.85\textwidth]{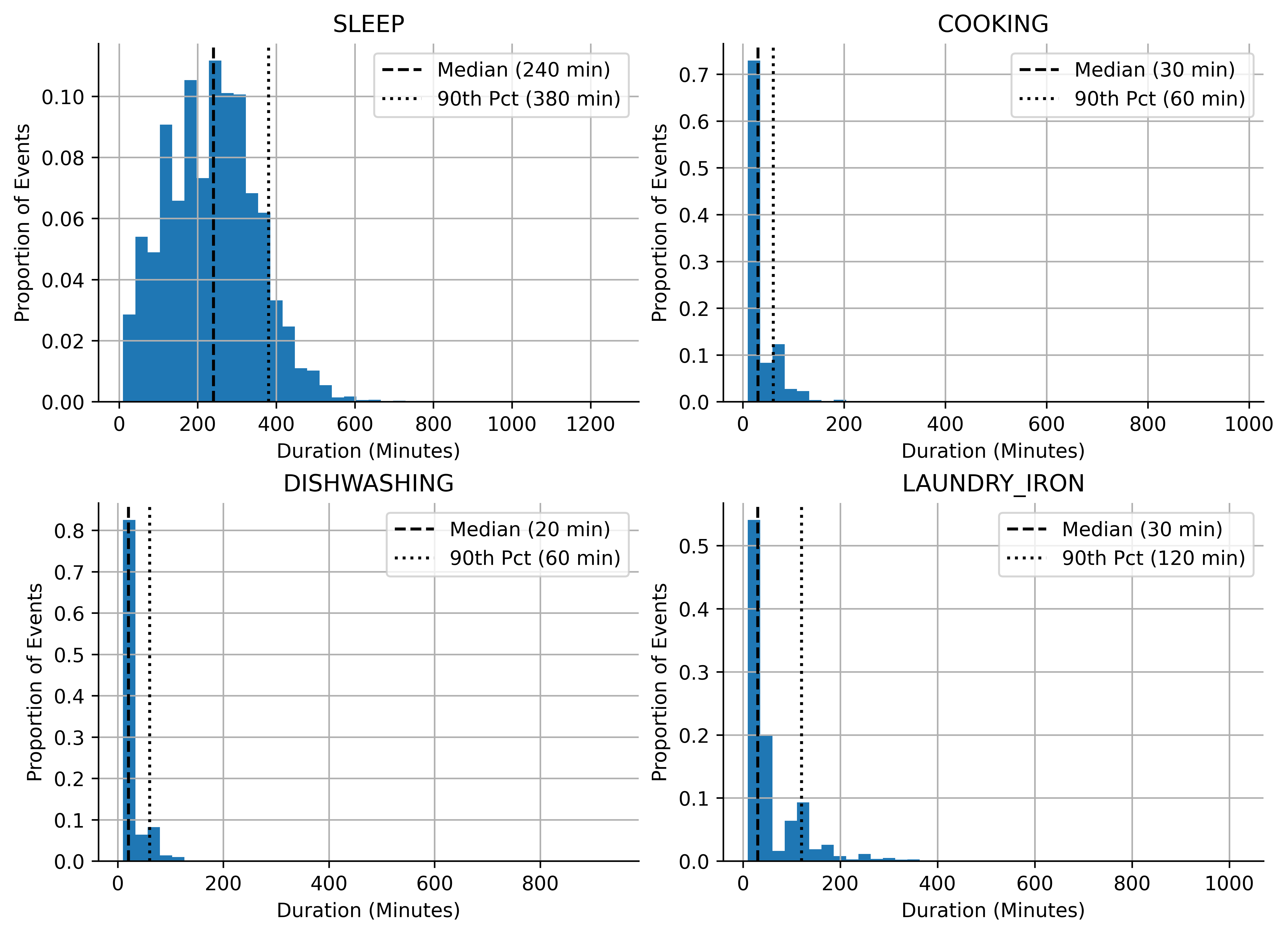} 
\caption{Empirical survival functions for key activities demonstrating that activity durations are not memoryless, motivating the use of a semi-Markov (hazard) model.}
\label{fig:dwell_curves}
\end{figure}

\paragraph{Justifying the Semi-Markov Approach.} A purely Markovian model assumes a constant probability of leaving that activity at any time step. Figure~\ref{fig:dwell_curves} shows the empirical dwell-time distributions for several key activities. For example, SLEEP shows a strong peak around 8 hours, and COOKING peaks around 20-30 minutes. The probability of leaving an activity is highly dependent on how long it has already been in progress. This non-constant hazard rate is a clear violation of the Markov assumption and provides a strong justification for employing a semi-Markov model.

\begin{figure}[!htbp]
\centering
\includegraphics[width=\textwidth]{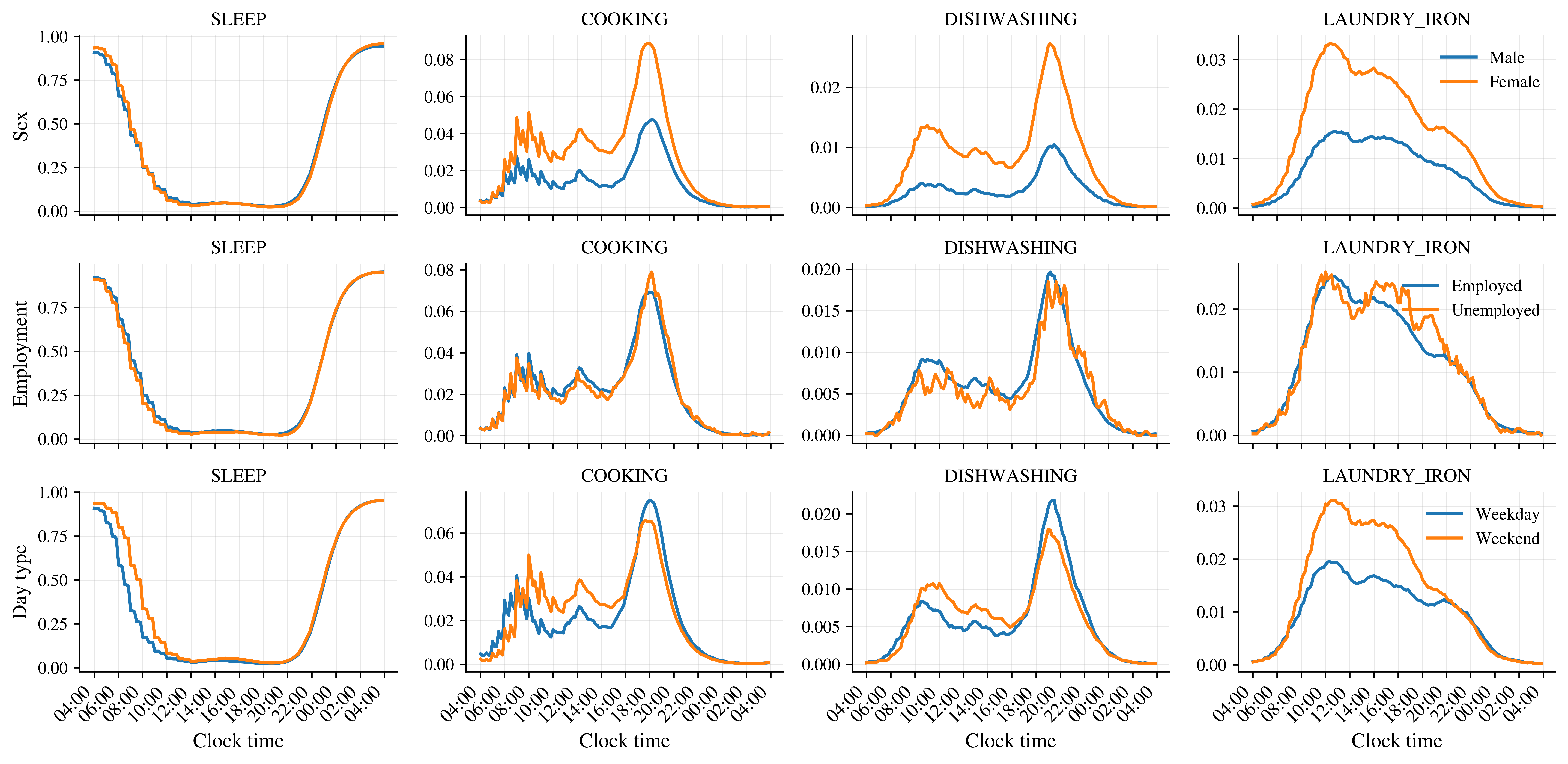}
\caption{Occupancy/activity probability curves for selected activities (\textsc{Sleep}, \textsc{Cooking}, \textsc{Dishwashing}, 
\textsc{Laundry/Ironing}), stratified by sex, employment status, and Day-Type. The strong separation between groups illustrates 
why these covariates drive significant predictive improvements.}
\label{fig:occ_covariates_compact}
\end{figure}

\paragraph{Covariate Activity Pattern Differences}
The importance of the covariates can be understood by visualizing their impact on daily activity patterns. Figure~\ref{fig:occ_covariates_compact} shows the probability of being engaged in selected activities (\textsc{Sleep}, \textsc{Cooking}, \textsc{Dishwashing}, \textsc{Laundry/Ironing}) over the course of the day, stratified by three key covariates: Sex, Employment Status, and Day-Type. The separations between the curves highlight why conditioning on these covariates may yield improvements in predictive accuracy.

\begin{figure}[h!]
\centering
\includegraphics[width=0.85\textwidth]{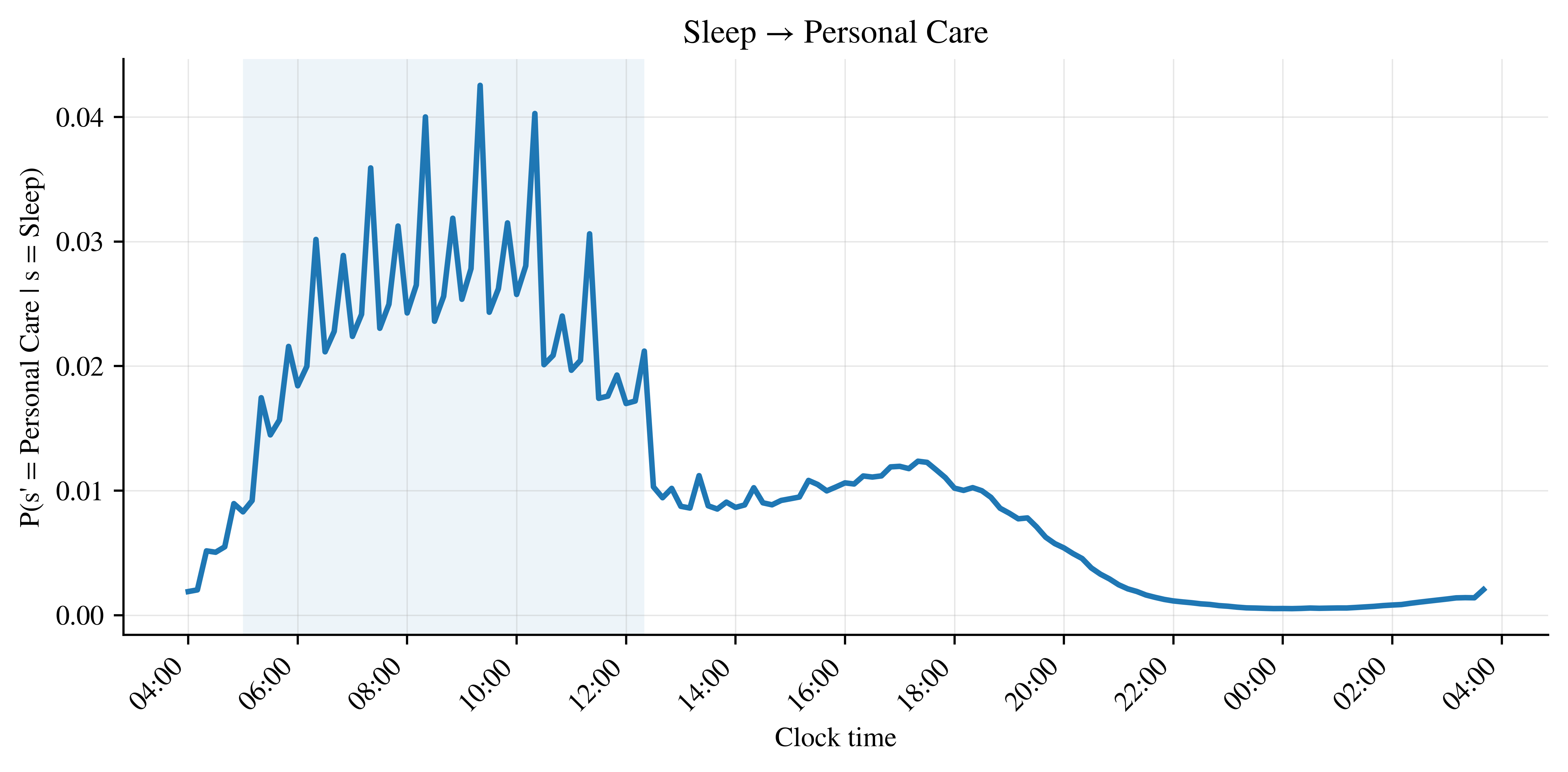} 
\caption{The empirical probability of transitioning from \textsc{Sleep} to \textsc{Personal Care} as a function of the time of day. The sharp morning peaks demonstrates the necessity of a time-inhomogeneous model.}
\label{fig:time_inhomogeneous}
\end{figure}

\paragraph{Justifying the Time-Inhomogeneous Approach.} A time-homogeneous model would assume that transition probabilities are constant throughout the day. The data strongly contradicts this assumption. Figure~\ref{fig:time_inhomogeneous} illustrates this by plotting the probability of transitioning from \textsc{Sleep} to \textsc{Personal Care} at each 10-minute interval. The probability is near zero for most of the day but exhibits a dramatic spike between 5 a.m. and 10 a.m. This strong diurnal pattern, which is present for most activity pairs, necessitates the use of a time-inhomogeneous model with distinct transition matrices for each time slot.

\subsection{Experimental Design}
Our analysis uses the full span of publicly available ATUS data from 2003 through 2024 comprising \(252,808\) respondents. The complete set of respondent diaries is partitioned into a training set (80\%) and a holdout test set (20\%). A random split rather than a chronological (last-20\% as test) split is used as our goal is to evaluate average-case sequence modeling under the population distribution (by season/quarter, day-type, and demographics), not to forecast across exogenous calendar shocks.

The splits are stratified by the covariates used, ensuring similar composition in train and test for each covariate type. Each respondent contributes exactly one diary, so there is no subject leakage across splits. The model parameters are estimated using only the training data, and all performance metrics are computed on the unseen holdout data to ensure a robust evaluation of predictive performance.

A central goal of this paper is to quantify the predictive value of different demographic factors and the effects of dwell time models. To achieve this, we systematically compare a series of models, each conditioned on different respondent attributes.

\paragraph{The Baseline Model (S0).} We first establish a baseline model, denoted S0, which is ungrouped. In this model, all respondents are treated as belonging to a single, homogeneous population. This model captures the average, population-level activity patterns but does not account for any demographic variation.

\paragraph{Single-Covariate Models (S1-S6).} We then train a series of models where respondents are partitioned into groups based on a single demographic covariate. These models, labeled S1 through S6, are detailed in Table~\ref{tab:covariates}. For each factor, a separate model is trained; for instance, the S2 model partitions the data into two groups (\(g_i \in \{\text{Male}, \text{Female}\}\)) and estimates distinct parameters for each.

\begin{table}[htbp]
\centering
\caption{Single-Covariate Models Used for Comparison.}
\label{tab:covariates}
\begin{tabular}{lll}
\hline
\textbf{Model ID} & \textbf{Covariate} & \textbf{Groups / Bins} \\
\hline
S1 & Region & Northeast; Midwest; South; West \\
S2 & Sex & Male; Female \\
S3 & Employment Status & Employed; Not employed; Not in work force \\
S4 & Day-Type & Weekday; Weekend \\
S5 & Household (HH)-Size & 1 member; 2 members; 3 members; 4+ members \\
S6 & Quarter/Season & Winter (Q1); Spring (Q2); Summer (Q3); Fall (Q4) \\
\hline
\end{tabular}
\end{table}

\paragraph{Model Comparisons.} Our analysis focuses on two key comparisons. First, we measure the improvement of each single-covariate model (S1-S6) relative to the ungrouped S0 baseline. This isolates the value of each demographic factor. Second, for every model configuration, we compare the performance of the purely Markovian router model against the full semi-Markov version that includes the hazard component. We denote these enhanced models with an ``-H" suffix (e.g., S0-H, S1-H). This allows us to quantify the additional predictive gain from explicitly modeling activity durations.

\subsection{Survey Weights and Model Inputs}
A critical feature of the ATUS dataset is the final person-day weight, \(w_i > 0\), provided by ATUS for each diary. These weights represent the number of people in the population that each respondent represents. They are useful for correcting biases in the sample that arise from differences in sampling probabilities and response rates across various demographic groups and days of the week. By applying these weights, we can produce estimates that are statistically representative of the entire target population. Thus, we use these weights throughout our analysis to ensure our findings are generalizable and reflect the true composition and behavior of the population.

In training, the statistics for the Markov model and dwell periods are computed as weighted sums. In evaluation, our metrics, such as negative log-likelihood, are also weighted. This ensures that our model's parameters and performance are representative of the target population, not the raw sample composition. The final inputs to train our mathematical model for each respondent \(i\) at each time slot \(t\) are the activity state \(S_{i,t}\), their demographic group \(g_i\), the current run-length of the activity \(\ell_{i,t}\), and the coarse day-part block \(b(t)\).

\section{Mathematical Model}\label{s:model}

We model daily activity sequences as a time-inhomogeneous semi-Markov process, discretized into \(T=144\) ten-minute slots. The mathematical notations for this framework are described in the Table \ref{tab:nomenclature}. This framework is explicitly designed to capture both the fine-grained, time-of-day variations in activity choice and the realistic durations of those activities. Our approach combines a time-inhomogeneous Markov router to model the sequence of activities with a hazard model to control the dwell time in each state.

\begin{table}[h!]
\centering
\caption{Nomenclature used throughout the model description.}
\label{tab:nomenclature}
\footnotesize
\setlength{\tabcolsep}{6pt}
\renewcommand{\arraystretch}{1.15}
\begin{tabularx}{\linewidth}{>{\raggedright\arraybackslash}p{0.28\linewidth} X}
\toprule
\textbf{Symbol} & \textbf{Description} \\
\midrule
$K$ & Number of activity states in the set $\mathcal{S}=\{1,\dots,K\}$. \\
$T$ & Number of 10-minute slots in a day ($T=144$). \\
$S_{i,t}$ & Activity state of respondent $i$ at slot $t$. \\
$g_i\in\mathcal{G}$ & Demographic group of respondent $i$. \\
$b(t)$ & Day-part block for time slot $t$. \\
$w_i$ & Survey design weight for respondent $i$. \\
$\ell_{i,t}$ & Run length (in slots) of the current activity at the start of slot $t$. \\
$\mathcal{L}=\{L_1,\dots,L_M\}$ & Set of disjoint run-length bins for discretizing duration. \\
$C_{s,s'}^{(g,t)}$ & Weighted count of transitions $s \to s'$ for group $g$ at slot $t$. \\
$C_{s,s'}^{(b)}$ & Total weighted transition count for $s \to s'$ within block $b$. \\
$N_{s,m}^{(g,t)}$ & Weighted count of exposures in run-length bin $L_m$ for group $g$ at slot $t$. \\
$E_{s,m}^{(g,t)}$ & Weighted count of exits from run-length bin $L_m$ for group $g$ at slot $t$. \\
$\bar{\theta}_{s,s'}^{(b)}$ & Block-level prototype probability for the transition $s \to s'$. \\
$\widehat{\theta}_{s,s'}^{(g,t)}$ & Posterior estimate of the router probability for $s \to s'$. \\
$\bar{h}_{s,m}^{(b)}$ & Block-level prototype hazard probability for run-length bin $L_m$. \\
$\widehat{h}_{s,m}^{(g,t)}$ & Posterior estimate of the hazard probability. \\
$\varphi_{s,s'}^{(g,t)}$ & Conditional destination probability of transitioning to $s'$. \\
$\tau_b,\,k$ & Hyperparameters for the router model, controlling shrinkage and smoothing. \\
$\kappa_b$ & Hyperparameter for the hazard model, controlling shrinkage. \\
\bottomrule
\end{tabularx}
\end{table}
The primary statistical challenge is data sparsity, magnified by the scale of our time-inhomogeneous model. For even a single demographic group, estimating the full set of transition probabilities requires fitting \(143\) distinct \(14 \times 14\) transition matrices that is one for each 10-minute slot of the day which amounts to estimating \(28,028\) parameters. When the data is further stratified by demographic covariates, the number of observations available for any specific group at a particular time of day becomes extremely thin, making direct estimation from counts highly unstable and prone to overfitting.

To address this, our framework uses a hierarchical approach. We regularize the slot-level estimates by pooling them toward robust, data-rich estimates computed over coarser day-part blocks. This method allows the model to borrow statistical strength across time, preventing overfitting in sparse cells while preserving the ability to capture specific, time-dependent patterns where the data permit.

\subsection{Day-Part Blocks for Hierarchical Shrinkage}
Our approach is the use of four coarse day-part blocks to create stable priors: \textbf{Night} (10:00 p.m.–5:50 a.m.), \textbf{Morning} (6:00 a.m.–11:50 a.m.), \textbf{Afternoon} (12:00 p.m.–5:50 p.m.), and \textbf{Evening} (6:00 p.m.–9:50 p.m.).

Within each block \(b\), we compute a single, low-variance prototype transition matrix, \(\bar{\Theta}^{(b)}\). This is achieved by first aggregating the weighted transition counts, \(C_{s,s'}^{(g,t)}\) (Appendix \ref{A:Count}), across all demographic groups \(g\) and all time slots \(t\) that fall within that block:
\[
C_{s,s'}^{(b)} = \sum_{g \in \mathcal{G}} \sum_{t \in b} C_{s,s'}^{(g,t)}.
\]
The prototype probability, \(\bar{\theta}_{s,s'}^{(b)}\), is then the simple row-normalization of these aggregated counts:
\begin{align}
\bar{\theta}_{s,s'}^{(b)} = \frac{C_{s,s'}^{(b)}}{\sum_{j=1}^{K} C_{s,j}^{(b)}}. \label{eq:prior}
\end{align}
This prototype represents the average transition behavior for that part of the day (e.g., a morning matrix) and serves as an informative prior for the models at each specific 10-minute slot. This shrinkage mechanism ensures that when 10-minute slot-level data is sparse, our estimates are pulled toward the reliable block-level average, preventing overfitting and increasing robustness.

\subsection{The Router Model (S0 -- S6)}
The router component defines the purely Markovian transition dynamics of our system. It determines the probability, \(\theta_{s,s'}^{(g,t)}\), of moving from activity \(s\) to \(s'\). We model each row of the transition matrix as a draw from a Dirichlet distribution whose parameters are set by the corresponding row of the block prototype, \(\bar{\theta}_{s,\cdot}^{(b(t))}\) from Equation (\ref{eq:prior}):
\[
\theta_{s,\cdot}^{(g,t)} \mid \bar{\theta}_{s,\cdot}^{(b(t))} \sim \mathrm{Dir}\!\Big( \tau_b\,\bar{\theta}_{s,\cdot}^{(b(t))} + k\,\tfrac{\mathbf{1}}{K} \Big).
\]
The design-weighted counts \(C_{s,s'}^{(g,t)}\) calculated by summing the survey weights of all respondents \(i\) in group \(g\) who transition from \(s\) to \(s'\) at time \(t\). The posterior mean for the transition probability is then:
\begin{align} \label{eq:slot_prob}
\widehat{\theta}_{s,s'}^{(g,t)}
=
\frac{ C_{s,s'}^{(g,t)} + \tau_b\,\bar{\theta}_{s,s'}^{(b(t))} + k/K }
     { \sum_{j=1}^K C_{s,j}^{(g,t)} + \tau_b + k }.
\end{align}
This formulation alone defines the family of purely Markovian models, from the ungrouped baseline (S0) to the single-covariate models (S1-S6). Here, \(\tau_b\) controls the degree of shrinkage towards the block-level prior, with larger values indicating stronger shrinkage. The hyperparameter \(k\) provides a small amount of smoothing to avoid zero probabilities for unobserved transitions.

\subsection{The Hazard Model for Durations}
To achieve realistic durations and define our semi-Markov models (the S-H series), we add a component that explicitly models the probability of an activity ending. For any given respondent, the decision to continue an activity or switch is a binary outcome. We model this using a discrete hazard probability, \(h_{s,m}^{(g,t)}\), which is the probability of \emph{leaving} state \(s\) during the current time slot, given the activity has lasted for a duration \(\ell_{i,t}\) falling in bin \(L_m\) which is defined as follows:

The run length \(\ell_{i,t}\) is simply how long the current activity has been going on when slot \(t\) starts,
measured in 10-minute steps (e.g., \(\ell_{i,t}=1\) means it just started; \(\ell_{i,t}=6\) means it has lasted for an hour).
Rather than estimate a separate leave-probability for every possible length, we group lengths into a few sensible ranges (``bins'') such as short, medium, and long. At each slot, the model looks up which bin \(\ell_{i,t}\) falls into and uses the corresponding parameter for that bin.

We use the Beta-Bernoulli conjugate model. For each cell \((s, m, g, t)\), the sufficient statistics are the weighted count of respondents \emph{exposed} (\(N\)) and the weighted count of those who \emph{exit} (\(E\)) as described in Appendix \ref{A:Count}. These are calculated by summing the survey weights \(w_i\) of the relevant respondents. We place a Beta prior on the hazard probability, centering it on the robust block-level prototype \(\bar{h}_{s,m}^{(b(t))}\):
\begin{align*}
h_{s,m}^{(g,t)} \mid \bar{h}_{s,m}^{(b(t))} \sim \mathrm{Beta}\!\Big( \kappa_b\,\bar{h}_{s,m}^{(b(t))}, \; \kappa_b(1-\bar{h}_{s,m}^{(b(t))}) \Big).
\end{align*}
The posterior mean for the hazard is a simple shrinkage estimator combining the local evidence with the block prior:
\begin{align} \label{eq: slot hazard}
\widehat{h}_{s,m}^{(g,t)}
=
\frac{ E_{s,m}^{(g,t)} + \kappa_b\,\bar{h}_{s,m}^{(b(t))} }
     { N_{s,m}^{(g,t)} + \kappa_b }.
\end{align}
The hyperparameter \(\kappa_b\) controls the degree of pooling, ensuring stable hazard estimates. The complete derivation of posteriors for the router and the hazard models is showed in the Appendix \ref{app:derivations}.

\subsection{The Combined Semi-Markov Process (S0-H -- S6-H)}

The router and hazard components from Equation (\ref{eq:slot_prob}) and \ref{eq: slot hazard} are synthesized to form the final transition probabilities for the semi-Markov models (S0-H through S6-H). For a respondent in state \(s\) at time \(t\), the model first decides whether to leave with probability \(\widehat{h}_{s,m}^{(g_i,t)}\). If a change occurs, it then decides where to go using the router's conditional destination distribution, \(\varphi_{s,s'}^{(g,t)}\):
\begin{align} \label{eq: cond prob}
\varphi_{s,s'}^{(g,t)}
=
\frac{\widehat{\theta}_{s,s'}^{(g,t)}}{ 1-\widehat{\theta}_{s,s}^{(g,t)} },\qquad \text{for } s'\neq s.
\end{align}
Using the conditional probability in Equation (\ref{eq: cond prob}) the complete one-step semi-Markov transition probability is thus:
\begin{align}
\Pr(S_{i,t+1}=s' \mid S_{i,t}=s,\, \ell_{i,t}\in L_m,\, g_i,\, t)
=
\begin{cases}
1-\widehat{h}_{s,m}^{(g_i,t)}, & \text{if } s'=s, \\[5pt]
\widehat{h}_{s,m}^{(g_i,t)}\,\varphi_{s,s'}^{(g_i,t)}, & \text{if } s'\neq s.
\end{cases}
\end{align}
To begin the generative process for a synthetic diary, the model must first draw an initial state, \(S_1\), for the first time slot of the day (4:00 a.m.). We estimate a separate initial state distribution, \(\pi^{(g)}\), for each demographic group \(g\). This distribution is estimated from the design-weighted frequencies of observed activities at \(t=1\). The detailed formulation for the estimate of this distribution is provided in Appendix~\ref{A:initial state}. The generative process for simulating a new 24-hour diary is detailed in Algorithm~\ref{alg:generation}.

\begin{algorithm}[!htpb]
\caption{Generative algorithm for simulating a 24-hour activity sequence}
\label{alg:generation}
\begin{algorithmic}[1]
\State \textbf{Input:} Demographic group \(g\), trained parameters 
       \(\{\widehat{\pi}^{(g)}, \widehat{\Theta}^{(g,t)}, \widehat{h}^{(g,t,m)}\}\)
\State \textbf{Output:} Synthetic activity sequence \((S_1, \dots, S_{144})\)
\Statex \textbf{Initialization}
\State Draw initial state \(S_1 \sim \widehat{\pi}^{(g)}\)
\State Set run length \(\ell \gets 1\)
\Statex \textbf{Main loop}
\For{$t = 1$ to $T-1$}
    \State Set current state \(s \gets S_t\)
    \State Identify run-length bin \(m\) such that \(\ell \in L_m\)
    \State Retrieve hazard probability \(\widehat{h}_{s,m}^{(g,t)}\)
    \State Draw \(u \sim U(0,1)\)
    \If{$u < \widehat{h}_{s,m}^{(g,t)}$} \Comment{Leave current state}
        \State Sample next state \(S_{t+1} \sim \varphi_{s,\cdot}^{(g,t)}\)
        \State Reset run length: \(\ell \gets 1\)
    \Else \Comment{Stay in current state}
        \State Set \(S_{t+1} \gets s\)
        \State Update run length: \(\ell \gets \ell + 1\)
    \EndIf
\EndFor
\State \textbf{return} \((S_1, \dots, S_{144})\)
\end{algorithmic}
\end{algorithm}
\section{Results}\label{s:results}

We evaluate our models on a hold-out test set of ATUS diaries to assess their predictive performance. Our analysis is designed to answer two primary questions: 
\begin{itemize}
    \item How much predictive power is gained by explicitly modeling activity durations using the hazard component? 
    \item Which demographic covariates provide the most significant improvements in predictive accuracy in both Markov and semi-Markov models?
\end{itemize}

\subsection{Evaluation Metrics}
The performance of each model is quantified on the held-out dataset using two distinct but complementary metrics. Let \(\mathcal{T}_i\) be the set of valid transition times for respondent \(i\), and let \(p(\cdot)\) be the one-step probability from the model.

\paragraph{Negative Log-Likelihood (NLL).}
The primary evaluation is based on the design-weighted NLL, a proper scoring rule averaged per transition. The NLL provides a rigorous measure of a model's predictive accuracy by calculating the average negative log-probability assigned to the true, observed sequence of activities. A lower NLL score indicates superior performance, as it reflects a higher likelihood assigned to the ground-truth data. As a sensitive metric that evaluates the entire predicted probability distribution, NLL is particularly effective for distinguishing between the subtle performance differences of well-calibrated probabilistic models.

The design-weighted NLL per transition is:
\[
\mathrm{NLL}
~=~
\frac{\sum_i w_i \sum_{t\in\mathcal{T}_i} \big(-\log p(S_{i,t+1} \mid S_{i,t}, \ell_{i,t}, g_i, t)\big)}
     {\sum_i w_i\,|\mathcal{T}_i|}
\]

\paragraph{Top-1 Next-Activity Accuracy.}
To provide a more intuitive measure of performance, we also report the design-weighted Top-1 accuracy. This metric calculates the percentage of 10-minute slots for which the model's single most likely prediction for the next activity was correct. While less sensitive than NLL, Top-1 accuracy offers a straightforward and easily interpretable measure of the model's practical utility in predicting the single most probable outcome at each step.

The design-weighted Top-1 next-activity accuracy is the weighted proportion of time steps where the model's most likely prediction matches the observed activity:

\[
\mathrm{Top1}
~=~
\frac{\sum_i w_i \sum_{t\in\mathcal{T}_i} \mathbf{1}\Big\{S_{i,t+1} = \arg\max_{s' \in \mathcal{S}} p(S_{i,t+1}=s' \mid S_{i,t}, \ell_{i,t}, g_i, t)\Big\}}
     {\sum_i w_i\,|\mathcal{T}_i|}
\]
Using these metrics, the following sections will present a systematic comparison of our model configurations. 

\subsection{Overall Model Performance} \label{ss:overall}
The primary findings of our study reveal the following patterns. First, adding the hazard component to explicitly model activity durations provides a substantial and consistent improvement in performance across all model configurations. Second, a hierarchy emerges among the demographic covariates, with 'Sex', ‘Day-Type‘ and ‘HH-Size‘ providing gains in predictive power. Table 3 presents the quantitative results on the hold-out set supporting these conclusions, comparing the NLL and Top-1 next-activity accuracy for the purely Markovian models (S0-S6) and their semi-Markov counterparts (S0-H to S6-H).

\begin{table}[!h]
\centering
\caption{Predictive Performance of All Models on the holdout set. Average one-step NLL is a measure of predictive error where lower is better. The $\Delta$NLL column shows the reduction in error relative to the corresponding baseline (S0 or S0-H). Top-1 Accuracy is the weighted percentage of correctly predicted next activities, where higher is better.}
\label{tab:model_results}
\begin{tabular}{lccc}
\hline
\textbf{Model} & \textbf{NLL} & \textbf{$\Delta$NLL (vs.S0 Baseline)} & \textbf{Top-1 Acc.} \\
\hline
\multicolumn{4}{c}{\textit{Markov Models (Router-Only)}} \\
\hline
S0 (Baseline)     & 0.438585 & --        & 90.2689\% \\
S1 (Region)       & 0.438554 & 0.00003   & 90.2689\% \\
S2 (Sex)          & 0.436974 & 0.00161   & 90.2689\% \\
S3 (Employment)   & 0.438605 & -0.00002  & 90.2689\% \\
S4 (Day-Type)     & 0.437812 & 0.00077   & 90.2689\% \\
S5 (HH-Size)      & 0.437615 & 0.00097   & 90.2689\% \\
S6 (Quarter/Season)       & 0.438580 & 0.00001   & 90.2689\% \\
\hline
\multicolumn{4}{c}{\textit{Semi-Markov Models (Router + Hazard)}} \\
\hline
S0-H (Baseline)     & 0.426483 & --        & 90.2733\% \\
S1-H (Region)       & 0.426554 & -0.00007  & 90.2730\% \\
S2-H (Sex)          & 0.424811 & 0.00167   & 90.2727\% \\
S3-H (Employment)   & 0.426566 & -0.00008  & 90.2714\% \\
S4-H (Day-Type)     & 0.425577 & 0.00091   & 90.2733\% \\
S5-H (HH-Size)      & 0.425396 & 0.00109   & 90.2739\% \\
S6-H (Quarter/Season)       & 0.426559 & -0.00008  & 90.2733\% \\
\hline
\end{tabular}
\vspace{0.25em}
\end{table}

\subsection{Bootstrap Confidence Intervals for Model Comparison}
To ensure that the observed differences in performance between models in Section \ref{ss:overall} are statistically meaningful and not merely due to the specific random sample of diaries in our holdout set, we employ a paired bootstrap procedure. The process is as follows: we generate 2,000 bootstrap replicates of the holdout set with replacement by resampling the respondents (diaries). For each replicate, we compute our metric of interest (e.g., the difference in NLL between model S0 and S0-H). This process yields an empirical distribution of the performance difference. We then calculate the 2.5th and 97.5th percentiles of this distribution to construct a 95\% confidence interval (CI). The interpretation of this interval is direct: if the 95\% CI for a performance difference does not contain zero, we can conclude with high confidence that the observed improvement is statistically significant.

Bootstrap analysis shows that the most significant performance gain across all models comes from adding the hazard component. Using the bootstrap analysis, we verify that this improvement is statistically robust and not a result of sampling variability. For each model configuration (S0 through S6), we compared the NLL of the purely Markovian model against its semi-Markov counterpart.

\begin{figure}[htb]
  \centering
  \includegraphics[width=\linewidth]{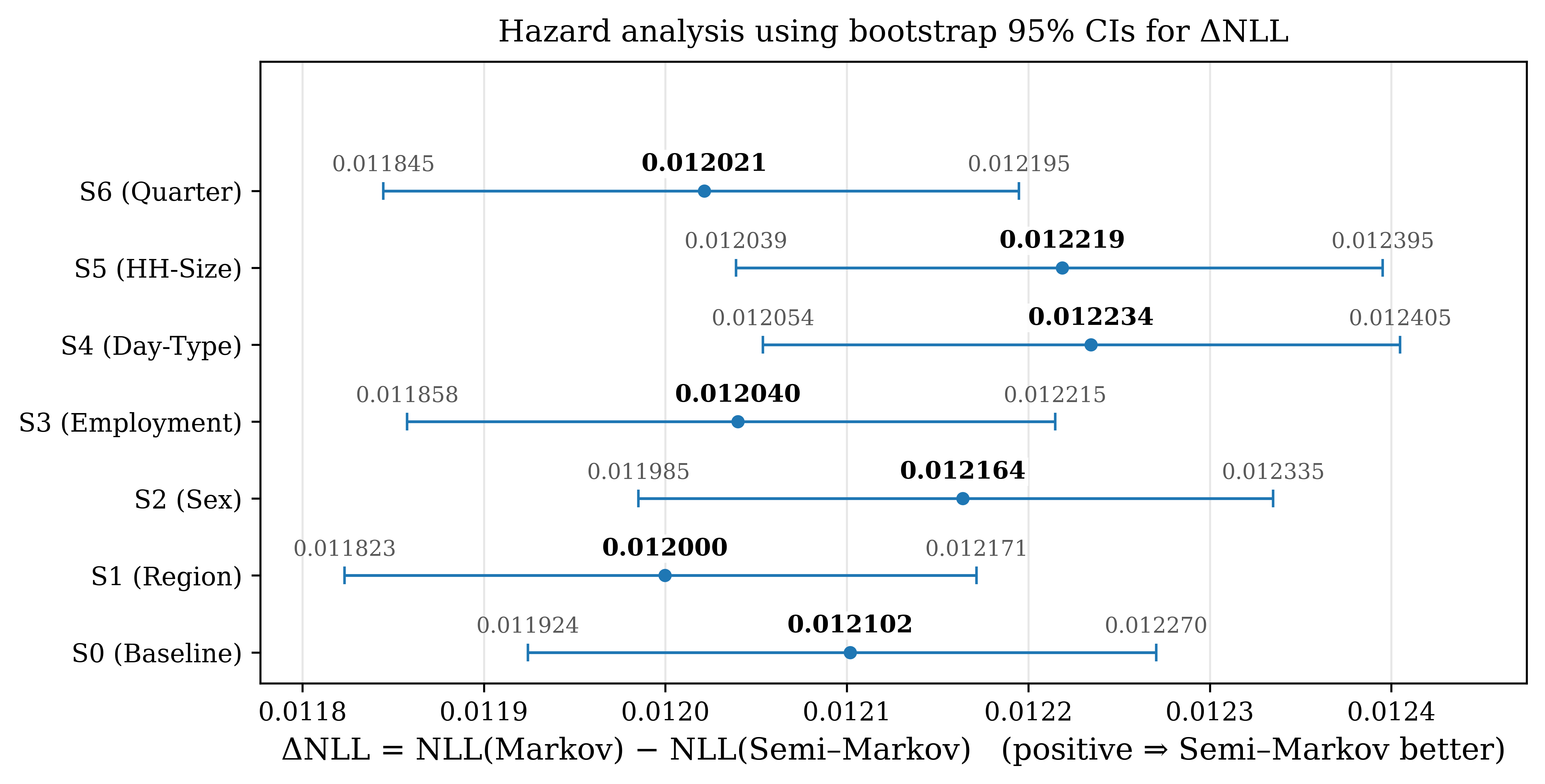}
  \caption{%
  \textbf{Hazard analysis using bootstrap 95\% CIs for} $\boldsymbol{\Delta\mathrm{NLL}}$.
  We plot $\Delta\mathrm{NLL}=\mathrm{NLL}(\text{Markov})-\mathrm{NLL}(\text{Semi\mbox{-}Markov})$, so positive values indicate that the Semi--Markov model achieves lower (better) NLL.
  Points mark the bootstrap mean; horizontal bars show the 95\% percentile confidence interval; S0 uses no covariate.
  All improvements are statistically significant at the 95\% confidence level as no intervals contain $0$.
  }
  \label{fig:delta-nll-forest}
\end{figure}

The results in Figure-\ref{fig:delta-nll-forest} provides further details on this comparison: explicitly modeling activity durations yields a large and statistically significant improvement in predictive accuracy. For the baseline S0 model, adding the hazard component (S0-H) reduces the NLL by 0.012102 on average, and the 95\% confidence interval for this improvement is tightly bound far from zero. This pattern holds for every covariate, confirming that the semi-Markov approach is consistently and significantly superior to the Markovian model for human level activity sequence modeling.

\subsection{Ranking the Predictive Power of Covariates}
The reduction in NLL for each single-covariate model relative to the S0/S0-H baseline is visualized in Figure~\ref{fig:nll_cov_stacked}. In both the Markov and semi‑Markov settings, covariates with positive $\Delta \text{NLL}$ reduce predictive error, while those near or to the left of zero offer little or no benefit:
\begin{figure}[!htbp]
  \centering
  \includegraphics[width=\textwidth]{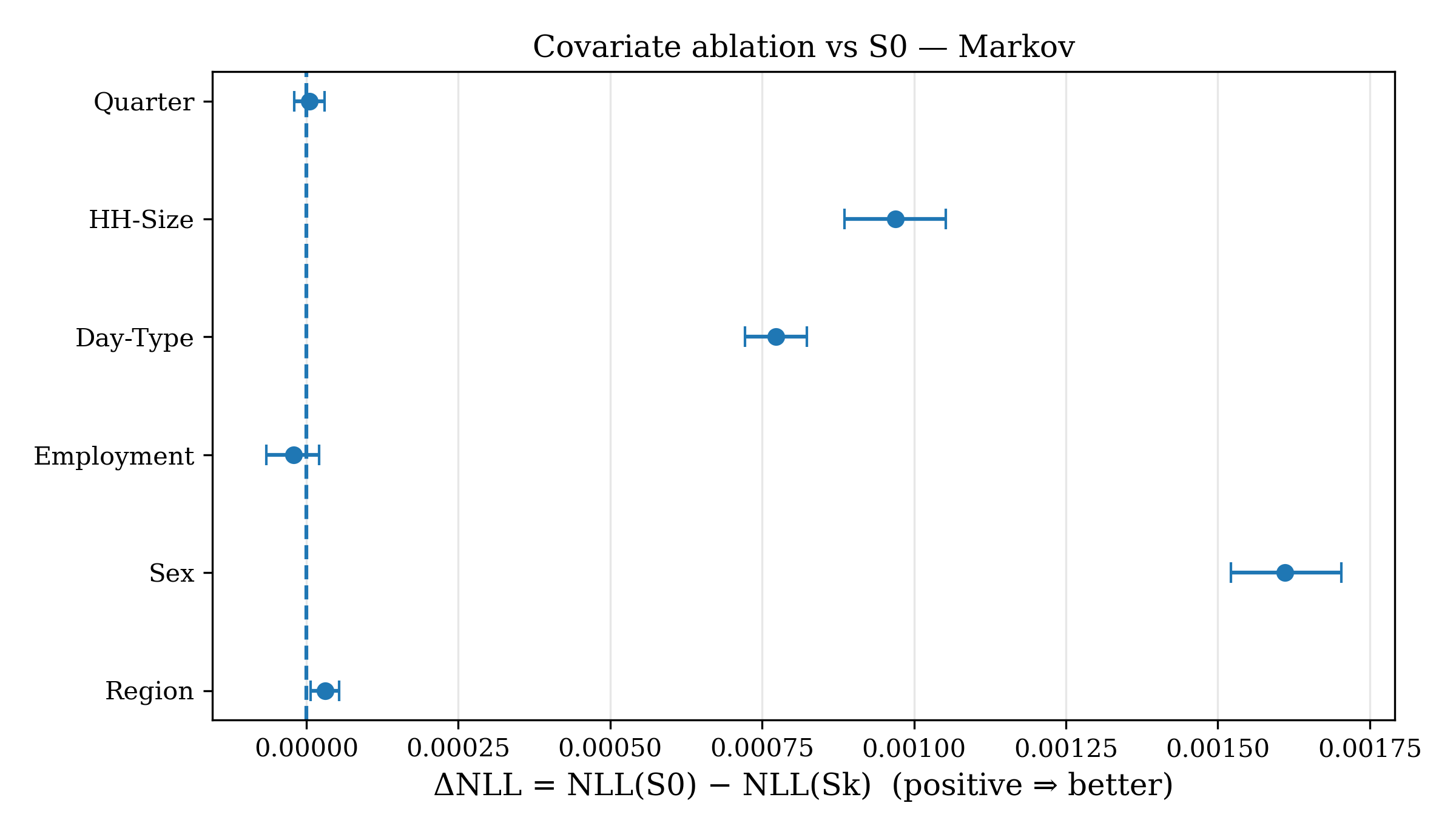}

  \vspace{0.6em}

hsmdad  \includegraphics[width=\textwidth]{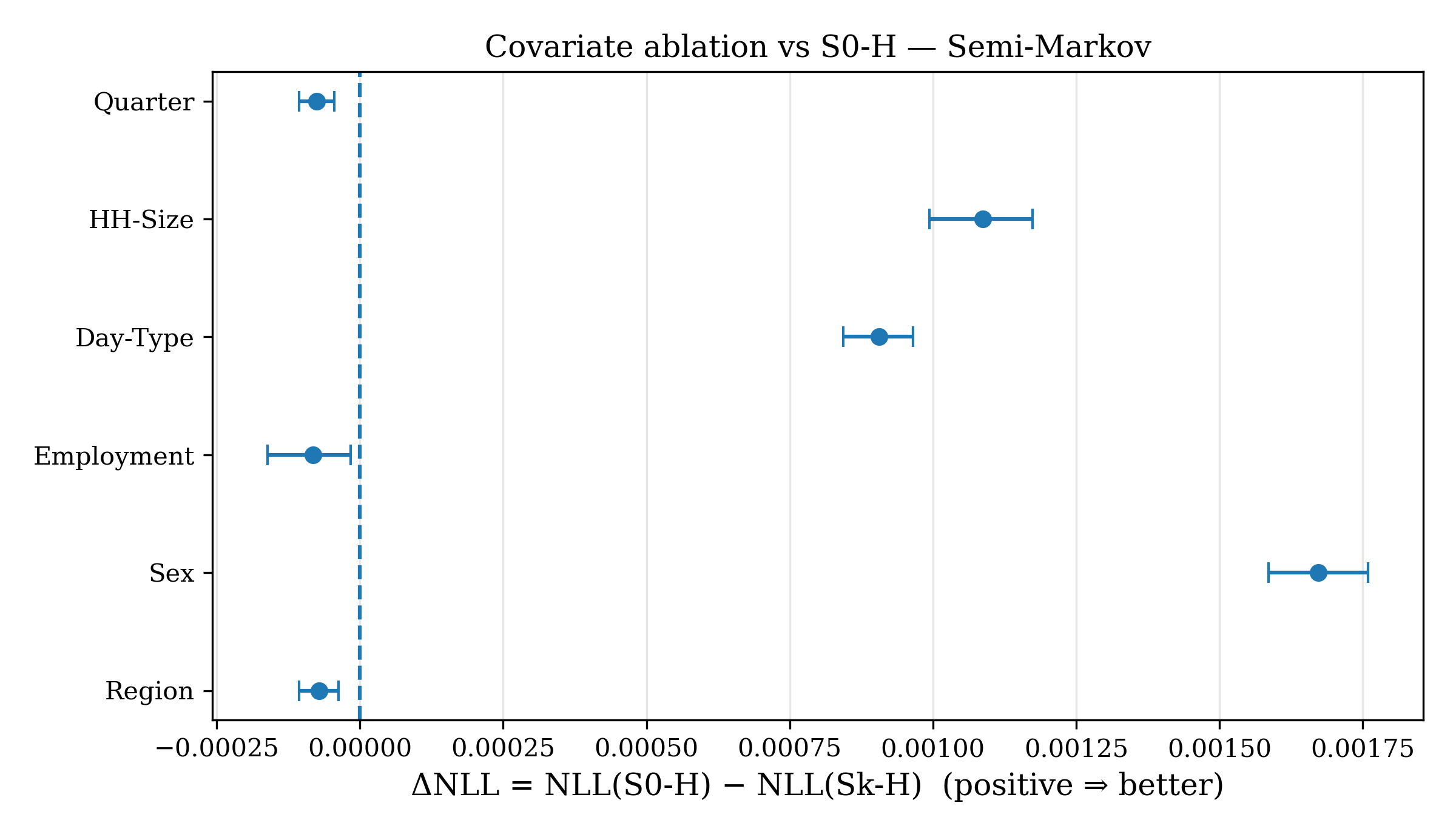}

  \caption{Bootstrap analysis for single-covariate ablations. Top: Markov (S1--S6 vs.\ S0). Bottom: Semi-Markov (S1-H--S6-H vs.\ S0-H).
  $\Delta \mathrm{NLL} = \mathrm{NLL}(\text{S0-H}) - \mathrm{NLL}(\text{Sk-H})$; Error bars: 95\% bootstrap CIs.}
  \label{fig:nll_cov_stacked}
\end{figure}
\begin{enumerate}
    \item \textbf{Sex (S2 / S2-H)} is the most influential covariate. 
    In both Markov and semi-Markov settings it yields the largest reduction in predictive error ($\Delta$NLL $\approx 0.0016$), 
    with confidence intervals well above zero, confirming robust predictive benefit.
    
    while in the semi-Markov model both remain beneficial and statistically significant.
    \item \textbf{HH-Size Band (S5 / S5-H)} and \textbf{Day-Type (S4 / S4-H)} provide the next largest gains. In the Markov model, HH-Size slightly outperforms Day-Type ($\Delta$NLL $\approx 0.0010$ vs.\ 0.0008), while in the semi-Markov model both remain beneficial and statistically significant.
    \item \textbf{Employment Status (S3 / S3-H)} does not significantly improve predictions. 
    Its estimates lie close to zero and confidence intervals overlap the baseline, 
    indicating little added value once durations are modelled.
    \item \textbf{Region (S1 / S1-H)} and \textbf{Quarter/Season (S6 / S6-H)} provide virtually no predictive gain for activity \textit{sequencing}.    
\end{enumerate}

Overall, the results show that Sex is the dominant covariate for next-activity prediction, with HH-Size and Day-Type providing complementary improvements. 
Other demographic factors such as employment, region, and season add little or no sequencing value once dwell times are incorporated.

\section{Discussion}\label{s:discussion}

Our results demonstrate that a hierarchical semi-Markov models can effectively capture the sequence and duration of human activities, and that a select few demographic factors are disproportionately important for this task. In this section, we interpret these findings, discuss their implications for energy systems modeling, and outline the limitations of our study.

\subsection{Interpretation of Key Findings}
Two central conclusions emerge from our analysis described as follows:

\paragraph{The Dominance of Duration Modeling.}
Adding dwell time for activities improved the models significantly. This finding underscores a fundamental limitation of memoryless property in Markov models for human behavior. Activities like sleeping or working have characteristic durations that are not geometrically distributed. By incorporating a run-length-dependent hazard function, the semi-Markov model aligns more closely with the temporal logic of human schedules, resulting in a substantial and statistically significant improvement in predictive accuracy.

\paragraph{The Hierarchy of Covariates.}
The results show that not all demographic factors are equally useful for predicting activity sequences. 
Sex emerges as the dominant driver of predictive improvement, reflecting systematic differences in daily routines followed by HH-Size and Day-Type which provided complementary gains, indicating that both family context and whether it is a weekday, weekend, or holiday meaningfully shape activity sequencing. 
By contrast, Employment Status offers little benefit once durations are accounted for, and Region and Season contribute negligibly—or even slightly degrade performance in the semi-Markov setting. 
This distinction is important; although Region and Season are critical for downstream energy calculations (e.g., HVAC loads), they do not appear to alter the underlying pattern of human behavior in terms of activity sequencing.

\subsection{On the Magnitude and Significance of Model Improvements}
While the per-transition NLL improvements may appear marginal (e.g., 0.012219 for S1-H vs. SH), it is crucial to recognize that the likelihood of a model generating an entire sequence is exponentially related to the total NLL. A small, consistent improvement over each of the 143 transitions in a day compounds into a substantial increase in the model's overall explanatory power.

An improvement of $\Delta\text{NLL}$ means the better model is \(e^{\Delta\text{NLL}}\) times more likely to predict the next step correctly. Over a full day, this effect compounds multiplicatively. We can see the significance of this by comparing the two main findings from our results:

First, consider the dominant effect of adding the hazard model. The improvement of the S1-H model over the S1 baseline was approximately 0.2094. For a full day's sequence, the likelihood ratio is:
\[
\text{Likelihood Ratio (Hazard)} = (e^{0.0.012219})^{143} = e^{1.747317} \approx 5.7391
\]
This number signifies that the semi-Markov model is more likely to have generated the observed sequences than the purely Markovian model, confirming the critical importance of modeling durations.

Now, consider the moderate improvement from adding the best covariate, Sex (S2-H vs. S0-H), which had a $\Delta\text{NLL}$ of 0.0016. The likelihood ratio for a full day is:
\[
\text{Likelihood Ratio (Covariate)} = (e^{0.0016})^{143} = e^{0.2288} \approx 1.2570
\]
This means that for a typical 24-hour diary, the model conditioned on covariate Sex is about $25\%$ more likely to have generated the observed sequence of activities than the ungrouped semi-Markov baseline. While a smaller figure, it represents a consistent and statistically significant improvement in explanatory power that originates from a seemingly small per-step gain.

\subsection{Implications for Energy Systems Modeling}
While this paper focuses on the prediction of activity sequences, its findings have direct practical implications for the design and operation of modern energy systems. Our model provides the high-fidelity behavioral inputs required for a range of downstream energy analysis tasks.

\paragraph{A Foundation for High-Fidelity Load Profile Generation.}
The primary downstream application of our model is to serve as a stochastic activity generator for bottom-up residential load profile simulations. By producing more realistic activity sequences, especially with respect to the timing and duration of high-consumption activities (e.g., \textsc{Cooking}, \textsc{Screens/Leisure}), our framework enables the generation of more accurate load profiles. These high-fidelity profiles are essential for tasks that depend on realistic peak demand characteristics, such as sizing neighborhood-scale battery storage, assessing grid stability, and evaluating the impact of new technologies.

\paragraph{Informing the Electrification Transition.}
Our model can be used to explore critical scenarios, such as the timing of Electric Vehicle (EV) charging. By identifying realistic and demographically-varying "at-home" windows, the model provides a data-driven basis for simulating charging patterns and assessing their potential grid impact, moving beyond simple assumptions about overnight charging.

\subsection{Limitations and Future Research}
While our framework demonstrates strong performance, we acknowledge several limitations that point toward avenues for future research. First, our analysis is currently limited to \textbf{single covariates}. Exploring models that capture interaction effects is an important next step. Second, the \textbf{14-state activity taxonomy}, while interpretable, aggregates all "Out-of-Home" activities. A more granular taxonomy could provide deeper insights, especially for modeling EV charging behavior. Finally, our model treats individuals as \textbf{independent agents}. Extending the model to a multi-agent framework that captures intra-household dynamics is a challenging but important direction for future work.

\section{Conclusion}\label{s:conclusion}

This paper developed and validated a hierarchical semi-Markov model for generating realistic, demographically-conditioned daily activity sequences from nationally representative time-use data. By addressing the key challenges of data sparsity and activity duration modeling, our framework provides a robust foundation for human activity sequencing.

Our empirical evaluation yielded two primary findings. First, explicitly modeling activity durations via the semi-Markov hazard component is not merely an incremental improvement but a critical one, offering a substantial and statistically significant increase in predictive accuracy over purely Markovian approaches. Second, we identified a clear hierarchy in the predictive power of some demographic factors providing meaningful complementary gains while others seem to be negligible.

The resulting model is an interpretable and computationally efficient tool for generating the high-fidelity behavioral inputs needed to understand the impacts of residential electricity consumption and to design more effective energy management strategies. Future work will focus on extending this framework to capture interaction effects between covariates and to model the correlated activities of multiple individuals within a household, further enhancing the realism and policy relevance of the generated sequences.

\if0\blind{
\section*{Acknowledgements}
\noindent
The authors gratefully acknowledge funding from Triad National Security LLC under the grant from the Department of Energy National Nuclear Security Administration (award no. 89233218CNA000001).
} \fi
\bibliographystyle{unsrt}
\spacingset{1}
\bibliography{IISE-Trans1}

\clearpage
\newpage

\section*{Appendix}
\appendix

\section{Count Data}\label{A:Count}

\paragraph{Weighted Counts, Exposures, and Exits at the Slot-Level.}
The weighted count of transitions from state \(s\) to \(s'\) for group \(g\) at time \(t\) is:
\[
C_{s,s'}^{(g,t)} = \sum_{i: g_i=g} w_i \cdot \mathbf{1}\{S_{i,t}=s \text{ and } S_{i,t+1}=s'\}
.\]
The weighted number of individuals in group \(g\) exposed to the risk of leaving state \(s\) at time \(t\) with a run-length in bin \(L_m\) is:
\[
N_{s,m}^{(g,t)} = \sum_{i: g_i=g} w_i \cdot \mathbf{1}\{S_{i,t}=s \text{ and } \ell_{i,t} \in L_m\}.
\]
The weighted number of individuals who exit under these conditions is:
\[
E_{s,m}^{(g,t)} = \sum_{i: g_i=g} w_i \cdot \mathbf{1}\{S_{i,t}=s, S_{i,t+1} \neq s, \text{ and } \ell_{i,t} \in L_m\}.
\]

\paragraph{Aggregation to the Block-Level.}
The block-level statistics, used to form the priors, are created by summing the slot-level statistics across all time slots \(t\) within a block \(b\) and across all demographic groups \(\mathcal{G}\).

The total weighted transition count for \(s \to s'\) within block \(b\) is:
\[
C_{s,s'}^{(b)} = \sum_{g \in \mathcal{G}} \sum_{t \in b} C_{s,s'}^{(g,t)}.
\]
Similarly, the total weighted exposures and exits for the hazard model within a block are:
\[
N_{s,m}^{(b)} = \sum_{g \in \mathcal{G}} \sum_{t \in b} N_{s,m}^{(g,t)} \qquad \text{and} \qquad E_{s,m}^{(b)} = \sum_{g \in \mathcal{G}} \sum_{t \in b} E_{s,m}^{(g,t)}.
\]
These aggregated counts are then used to calculate the block-level prototype probabilities \(\bar{\theta}^{(b)}\) and \(\bar{h}^{(b)}\).

\section{Derivation of Posterior Means}\label{app:derivations}

This appendix provides the derivations for the posterior mean estimators used in the main paper for the router, hazard, and initial state models. The conjugate priors are leveraged to obtain simple, closed-form expressions for the posterior distributions.

\subsection{Router Model Posterior}

The router model estimates the transition probabilities \(\theta_{s,\cdot}^{(g,t)}\) from a state \(s\) to all other states \(s'\) for a specific group \(g\) at a time slot \(t\). This is a multiclass classification problem, for which the Dirichlet-Multinomial conjugacy is well-suited.

\begin{itemize}
    \item \textbf{Likelihood}: The observed data for transitions out of state \(s\) are the design-weighted counts \(C_{s,s'}^{(g,t)}\) for each destination state \(s'\). The likelihood of observing these counts, given the transition probabilities \(\theta_{s,\cdot}^{(g,t)}\), follows a Multinomial distribution:
    \[
    p(C_{s,1}^{(g,t)}, \dots, C_{s,K}^{(g,t)} \mid \theta_{s,\cdot}^{(g,t)}) \propto \prod_{s'=1}^{K} \left(\theta_{s,s'}^{(g,t)}\right)^{C_{s,s'}^{(g,t)}}.
    \]

    \item \textbf{Prior}: As specified in the paper, we place a Dirichlet prior on the vector of transition probabilities \(\theta_{s,\cdot}^{(g,t)}\). This prior is informed by the block-level prototype \(\bar{\theta}_{s,\cdot}^{(b(t))}\) and smoothed with a small constant \(k\). The prior is:
    \[
    \theta_{s,\cdot}^{(g,t)} \sim \mathrm{Dir}(\alpha_1, \dots, \alpha_K),
    \]
    where the concentration parameters are \(\alpha_{s'} = \tau_b\,\bar{\theta}_{s,s'}^{(b(t))} + k/K\). The probability density function is:
    \[
    p(\theta_{s,\cdot}^{(g,t)}) \propto \prod_{s'=1}^{K} \left(\theta_{s,s'}^{(g,t)}\right)^{\alpha_{s'} - 1}.
    \]

    \item \textbf{Posterior}: Due to the conjugacy of the Dirichlet prior and Multinomial likelihood, the posterior distribution of \(\theta_{s,\cdot}^{(g,t)}\) is also a Dirichlet distribution. The posterior is found by multiplying the prior and the likelihood:
    \begin{align*}
    p(\theta_{s,\cdot}^{(g,t)} \mid \text{data}) &\propto p(\text{data} \mid \theta_{s,\cdot}^{(g,t)}) \cdot p(\theta_{s,\cdot}^{(g,t)}), \\
    &\propto \left( \prod_{s'=1}^{K} (\theta_{s,s'}^{(g,t)})^{C_{s,s'}^{(g,t)}} \right) \cdot \left( \prod_{s'=1}^{K} (\theta_{s,s'}^{(g,t)})^{\alpha_{s'} - 1} \right), \\
    &\propto \prod_{s'=1}^{K} (\theta_{s,s'}^{(g,t)})^{C_{s,s'}^{(g,t)} + \alpha_{s'} - 1}.
    \end{align*}
    This is the kernel of a new Dirichlet distribution, \(\mathrm{Dir}(\alpha'_1, \dots, \alpha'_K)\), with updated parameters:
    \[
    \alpha'_{s'} = C_{s,s'}^{(g,t)} + \alpha_{s'} = C_{s,s'}^{(g,t)} + \tau_b\,\bar{\theta}_{s,s'}^{(b(t))} + k/K
    \]

    \item \textbf{Posterior Mean}: The mean of a Dirichlet distribution \(\mathrm{Dir}(\alpha'_1, \dots, \alpha'_K)\) is given by \(\mathbb{E}[\theta_{s,s'}] = \frac{\alpha'_{s'}}{\sum_{j=1}^K \alpha'_{j}}\). Using this, we get the posterior mean estimate \(\widehat{\theta}_{s,s'}^{(g,t)}\):
    \begin{align*}
    \widehat{\theta}_{s,s'}^{(g,t)} &= \frac{C_{s,s'}^{(g,t)} + \tau_b\,\bar{\theta}_{s,s'}^{(b(t))} + k/K}{\sum_{j=1}^K (C_{s,j}^{(g,t)} + \tau_b\,\bar{\theta}_{s,j}^{(b(t))} + k/K)}, \\
    &= \frac{C_{s,s'}^{(g,t)} + \tau_b\,\bar{\theta}_{s,s'}^{(b(t))} + k/K}{\left(\sum_{j=1}^K C_{s,j}^{(g,t)}\right) + \tau_b \left(\sum_{j=1}^K \bar{\theta}_{s,j}^{(b(t))}\right) + \left(\sum_{j=1}^K k/K\right)}.
    \end{align*}
    Since \(\sum_{j=1}^K \bar{\theta}_{s,j}^{(b(t))} = 1\), the denominator simplifies, yielding the expression in the paper:
    \[
    \widehat{\theta}_{s,s'}^{(g,t)} = \frac{ C_{s,s'}^{(g,t)} + \tau_b\,\bar{\theta}_{s,s'}^{(b(t))} + k/K }{ \sum_{j=1}^K C_{s,j}^{(g,t)} + \tau_b + k }.
    \]
\end{itemize}

\subsection{Hazard Model Posterior}
The hazard model estimates the probability \(h_{s,m}^{(g,t)}\) of leaving a state \(s\), given a run-length in bin \(L_m\). This is a binary outcome (leave vs. stay), making the Beta-Bernoulli conjugate model the natural choice.
\begin{itemize}
    \item \textbf{Likelihood}: The data consists of \(E_{s,m}^{(g,t)}\) weighted exits and \(N_{s,m}^{(g,t)} - E_{s,m}^{(g,t)}\) weighted stays, out of \(N_{s,m}^{(g,t)}\) total exposures. The likelihood of this outcome follows a Bernoulli process, which for aggregated counts is a Binomial distribution:
    \[
    p(E_{s,m}^{(g,t)} \mid h_{s,m}^{(g,t)}) \propto \left(h_{s,m}^{(g,t)}\right)^{E_{s,m}^{(g,t)}} \left(1-h_{s,m}^{(g,t)}\right)^{N_{s,m}^{(g,t)} - E_{s,m}^{(g,t)}}.
    \]

    \item \textbf{Prior}:A Beta prior is placed on the hazard probability \(h_{s,m}^{(g,t)}\), centered on the block-level prototype \(\bar{h}_{s,m}^{(b(t))}\). The Beta distribution is defined by two parameters, \(\alpha\) and \(\beta\):
    \[
    h_{s,m}^{(g,t)} \sim \mathrm{Beta}(\alpha, \beta),
    \]
    where \(\alpha = \kappa_b\,\bar{h}_{s,m}^{(b(t))}\) and \(\beta = \kappa_b(1-\bar{h}_{s,m}^{(b(t))})\). The parameter \(\kappa_b\) acts as a pseudo-count, controlling the strength of the prior.

    \item \textbf{Posterior}: The Beta prior is conjugate to the Bernoulli/Binomial likelihood. The posterior distribution is therefore also a Beta distribution with updated parameters \(\alpha'\) and \(\beta'\):
    \begin{align*}
    \alpha' &= \alpha + (\text{number of successes}) = \kappa_b\,\bar{h}_{s,m}^{(b(t))} + E_{s,m}^{(g,t)}, \\
    \beta' &= \beta + (\text{number of failures}) = \kappa_b(1-\bar{h}_{s,m}^{(b(t))}) + (N_{s,m}^{(g,t)} - E_{s,m}^{(g,t)}).
    \end{align*}

    \item \textbf{Posterior Mean}: The mean of a Beta distribution \(\mathrm{Beta}(\alpha', \beta')\) is \(\frac{\alpha'}{\alpha'+\beta'}\). Substituting our posterior parameters:
    \begin{align*}
    \widehat{h}_{s,m}^{(g,t)} &= \frac{\kappa_b\,\bar{h}_{s,m}^{(b(t))} + E_{s,m}^{(g,t)}}{\left(\kappa_b\,\bar{h}_{s,m}^{(b(t))} + E_{s,m}^{(g,t)}\right) + \left(\kappa_b(1-\bar{h}_{s,m}^{(b(t))}) + N_{s,m}^{(g,t)} - E_{s,m}^{(g,t)}\right)}, \\
    &= \frac{E_{s,m}^{(g,t)} + \kappa_b\,\bar{h}_{s,m}^{(b(t))}}{E_{s,m}^{(g,t)} + \kappa_b\,\bar{h}_{s,m}^{(b(t))} + \kappa_b - \kappa_b\,\bar{h}_{s,m}^{(b(t))} + N_{s,m}^{(g,t)} - E_{s,m}^{(g,t)}}, \\
    &= \frac{ E_{s,m}^{(g,t)} + \kappa_b\,\bar{h}_{s,m}^{(b(t))} }{ N_{s,m}^{(g,t)} + \kappa_b }.
    \end{align*}
    This is the shrinkage estimator presented in the main paper.
\end{itemize}

\subsection{Initial State Distribution Posterior} \label{A:initial state}
The derivation for the initial state distribution \(\widehat{\pi}^{(g)}\) is nearly identical to the router model, but simpler as it lacks the hierarchical prior from the block-level.
\begin{itemize}
    \item \textbf{Likelihood}: The data are the weighted counts of respondents in group \(g\) starting their day in state \(s\), which we denote \(C_s^{(g)} = \sum_{i:g_i=g} w_i \cdot \mathbf{1}\{S_{i,1}=s\}\). The likelihood follows a Multinomial distribution:
    \[
    p(C_1^{(g)}, \dots, C_K^{(g)} \mid \pi^{(g)}) \propto \prod_{s=1}^{K} \left(\pi_s^{(g)}\right)^{C_s^{(g)}}.
    \]

    \item \textbf{Prior}: A symmetric Dirichlet prior is used for smoothing, which corresponds to adding a small pseudo-count to each category.
    \[
    \pi^{(g)} \sim \mathrm{Dir}(k/K, \dots, k/K).
    \]

    \item \textbf{Posterior}: The posterior is a Dirichlet distribution with parameters updated by the observed counts:
    \[
    \pi^{(g)} \mid \text{data} \sim \mathrm{Dir}(C_1^{(g)} + k/K, \dots, C_K^{(g)} + k/K).
    \]

    \item \textbf{Posterior Mean}: The posterior mean is:
    \[
    \widehat{\pi}_s^{(g)} = \frac{C_s^{(g)} + k/K}{\sum_{j=1}^K (C_j^{(g)} + k/K)} = \frac{ \sum_{i:g_i=g} w_i \cdot \mathbf{1}\{S_{i,1}=s\} + k/K }{ \left( \sum_{i:g_i=g} w_i \right) + k }.
    \]
    This matches the expression in Appendix A.
\end{itemize}

\section{ATUS Data Columns}\label{app:data cols}

Table~\ref{tab:atus_columns} lists the ATUS microdata variables (columns) that were used in constructing our dataset. These include respondent identifiers, demographic attributes, survey weights, and activity-level details.

\begin{table}[h!]
\centering
\caption{ATUS data columns.}
\label{tab:atus_columns}
\begin{tabular}{p{3cm} p{11cm}}
\toprule
\textbf{Column Name} & \textbf{Description} \\
\midrule
\texttt{TUCASEID} & Unique ATUS household identifier for each respondent. \\
\texttt{TULINENO} & Line number identifying the respondent within the household. \\
\texttt{TUDIARYDATE} & Date of the diary day (used to derive weekday/weekend and quarter). \\
\texttt{TUFINLWGT} & Final person-day survey weight for population inference. \\
\texttt{TELFS} & Labor force status code (employment classification). \\
\texttt{TESEX} & Respondent’s sex (male/female). \\
\texttt{TEAGE} & Respondent’s age in years. \\
\texttt{TUACTIVITY\_N} & Activity sequence number for each diary episode. \\
\texttt{TUACTDUR24} & Duration (minutes) of each reported ATUS activity. \\
\texttt{TUTIER1CODE} & First two digits of the 6-digit activity code (major activity category). \\
\texttt{TUTIER2CODE} & Middle two digits of the 6-digit activity code (subcategory). \\
\texttt{TUTIER3CODE} & Last two digits of the 6-digit activity code (fine-grained detail). \\
\texttt{HH\_SIZE} & Household size (constructed from roster file). \\
\bottomrule
\end{tabular}
\end{table}

\section*{Code and Data Availability}

The source code for the models, data preprocessing scripts, model weights and matrices, and analysis notebooks presented in this work are publicly available on GitHub at: \url{https://github.com/Rohitd922/atus_analysis}. The ATUS microdata used in this study are available directly from the U.S. Bureau of Labor Statistics.
\end{document}